\documentclass[twoside]{article}
\usepackage{fleqn,espcrc2}
\usepackage{epsfig}
\usepackage{graphicx}
\usepackage[figuresright]{rotating}

\def\be{ \begin{equation} }
\def\bea{\begin{eqnarray}}
\def\eea{\end{eqnarray}}
\def\ee{ \end{equation} }

\def\d0d0{ D^0\bar{D}^0 }
\def\p0p0{ P^0\bar{P}^0 }
\def\qp2{ \Bigl| \frac{q}{p} \Bigr|^2 }
\def\pq2{ \Bigl| \frac{p}{q} \Bigr|^2 }
\def\rarr{ \rightarrow }

\newcommand{\AmS}{{\protect\the\textfont2
  A\kern-.1667em\lower.5ex\hbox{M}\kern-.125emS}}

\title{New FOCUS results on charm mixing and CP violation}
\author{Stefano Bianco
\thanks{on behalf of the FOCUS Collaboration. 
Coauthors are: J.M. Link, V.S. Paolone, M. Reyes, P.M. Yager
({\bf UC DAVIS}); J.C. Anjos,
 I. Bediaga, C. G\"obel, J. Magnin, J.M. de Miranda, I.M. Pepe, A.C. dos Reis,
 F. Sim\~ao ({\bf CPBF, Rio de Janeiro});
S. Carrillo, E. Casimiro, H. Mendez, \hbox{A.S\'anchez-Hern\'andez,},
 C. Uribe, F. Vasquez ({\bf CINVESTAV, M\'exico City});
L. Cinquini, J.P. Cumalat, J.E. Ramirez, B. O'Reilly, E.W. Vaandering ({\bf CU
        Boulder});
J.N. Butler, H.W.K. Cheung, I. Gaines, P.H. Garbincius, L.A. Garren,
    E. Gottschalk,     S.A. Gourlay, P.H. Kasper,
A.E. Kreymer, R. Kutschke ({\bf Fermilab}); S. Bianco, F.L. Fabbri, S. Sarwar,
A. Zallo ({\bf INFN Frascati}); C. Cawlfield, D.Y. Kim,
        K.S. Park, A. Rahimi,
J. Wiss ({\bf UI Champaign}); R. Gardner ({\bf Indiana }); Y.S. Chung,
J.S. Kang, B.R. Ko, J.W. Kwak,
K.B. Lee, S.S. Myung, H. Park ({\bf Korea University, Seoul}); G. Alimonti,
        M. Boschini, D. Brambilla,
B. Caccianiga, A. Calandrino, P. D'Angelo, M. DiCorato, P. Dini, M. Giammarchi,
        P. Inzani,
F. Leveraro, S. Malvezzi, D. Menasce, M. Mezzadri, L. Milazzo, L. Moroni,
    D. Pedrini,     F. Prelz, M. Rovere, A. Sala,
S. Sala ({\bf INFN and Milano}); T.F. Davenport III ({\bf UNC Asheville});
        V. Arena,
G. Boca, G. Bonomi, G. Gianini, G. Liguori, M. Merlo, D. Pantea,
        S.P. Ratti, C. Riccardi,
 P. Torre, L. Viola, P. Vitulo ({\bf INFN and Pavia});
H. Hernandez, A.M. Lopez, L. Mendez,
A. Mirles, E. Montiel, D. Olaya, J. Quinones, C. Rivera, Y. Zhang ({\bf
Mayaguez, Puerto Rico});
N. Copty, M. Purohit, J.R. Wilson ({\bf USC Columbia});
K. Cho, T. Handler ({\bf UT Knoxville}); D. Engh, W.E. Johns, M. Hosack,
M.S. Nehring, M. Sales, P.D. Sheldon,
K. Stenson, M.S. Webster ({\bf Vanderbilt}); M. Sheaff ({\bf Wisconsin,
Madison}); Y. Kwon ({\bf Yonsei University, Korea}).
} \\
 Laboratori Nazionali di Frascati dell'INFN -  via E.~Fermi 40, Frascati
I-00044  }
       
\begin{document}

\begin{abstract}
We present a summary of recent results on CP violation and mixing in the
charm quark sector based on a high-statistics sample collected by
photoproduction experiment FOCUS (E831 at Fermilab). We have measured the
difference in 
lifetimes for the $D^0$ decays: $D^0 \rightarrow K^-\pi^+$ and $D^0
\rightarrow K^-K^+$. This translates into a measurement of the $y_{CP}$
mixing parameter in the  $\d0d0$ system, under the assumptions that
$K^-K^+$ is an equal mixture of CP odd and CP even eigenstates, and CP
violation is negligible in the neutral charm meson system. We verified 
the latter assumption by searching for CP violating  asymmetry in the
Cabibbo suppressed decay modes 
$D^+ \to K^-K^+\pi^+$, $D^0 \to K^-K^+$ and $D^0 \to \pi^-\pi^+$.  
We report preliminary results on a measurement of the branching ratio
$\Gamma(D^{*+}\rightarrow \pi^+ (K^+\pi^-))/\Gamma(D^{*+}\rightarrow \pi^+
(K^-\pi^+))$. 
\vspace{1pc}
\end{abstract}

\maketitle

\section{INTRODUCTION}
Particle-antiparticle mixing in the charm sector has distinctive features
that make it a high sensitivity probe to search for New Physics.
Recently, the possibility of collecting large, high-quality samples of
fully reconstructed $D$ meson decays have germinated several new results 
\cite{exprev}. In \S~2 we review the basic formalism  of charm mixing and
CP violation (CPV), as well as the experimental techniques for  measuring
the mixing parameters. After briefly describing the FOCUS detector in
\S~3, in \S~4 we report on a new measurement of 
the difference in 
lifetimes for the $D^0$ decays: $D^0 \rightarrow K^-\pi^+$ and $D^0
\rightarrow K^-K^+$. In \S~5 we show how we searched for  CPV asymmetry in the
Cabibbo suppressed decay modes 
$D^+ \to K^-K^+\pi^+$, $D^0 \to K^-K^+$ and $D^0 \to \pi^-\pi^+$.  Finally,
in \S~6 we report preliminary results on a measurement of the branching ratio
$\Gamma(D^{*+}\rightarrow \pi^+ (K^+\pi^-))/\Gamma(D^{*+}\rightarrow \pi^+
(K^-\pi^+))$. 
\section{CHARM MIXING AND CPV}
 Let us  recall the key features  of particle-antiparticle mixing
 \cite{Bigi:2000yz}\cite{Branco:1999fs}\cite{Leader96}.
 Because of weak interactions, flavor $f=s,c,b$ of a generic pseudoscalar
 neutral meson $P^0$ is not conserved. Therefore, it will try and decay
 with new mass eigenstates  $P^0_{1,2}$ which no longer carry definite
 flavor ${f}$: they are new  states with different mass and lifetime
 $| P^0_{1,2} \rangle  \propto  ( p|P^0 \rangle \pm
 q|\bar{P}^0\rangle) $
 where complex parameters $p$ and $q$ account for any CPV.
 The time evolution of $|P^0(t)\rangle$ is given by the
 Schr\"odinger equation. After a time $t$ the probability of finding the
 state $P^0$ transformed into $\bar{P}^0$ is
\bea
 |\langle \bar{P}^0|P^0(t)\rangle|^2 & \propto &  \qp2 e^{-\Gamma_1 t}
  [1+e^{\Delta \Gamma t} + \nonumber \\
    & & 2e^{\frac{\Delta\Gamma}{2}t} \cos (\Delta m t)] 
\eea
 with definitions  $\Delta m \equiv m_1-m_2$, $\Delta \Gamma = \Gamma_1
 -\Gamma_2$ and $\bar{\Gamma} \equiv (\Gamma_1+\Gamma_2)/2$.
 The two states will oscillate with  a rate  expressed by $\Delta
 m$ and $\Delta \Gamma$, which are usually calibrated
 by the  average decay rate by means of the mixing parameters
\begin{equation}
 x \equiv \Delta m/\bar{\Gamma} \qquad  
 y \equiv \Delta\Gamma/(2\bar{\Gamma})
\end{equation}
\begin{figure}[t]
 \vspace{2.5cm}
 \includegraphics{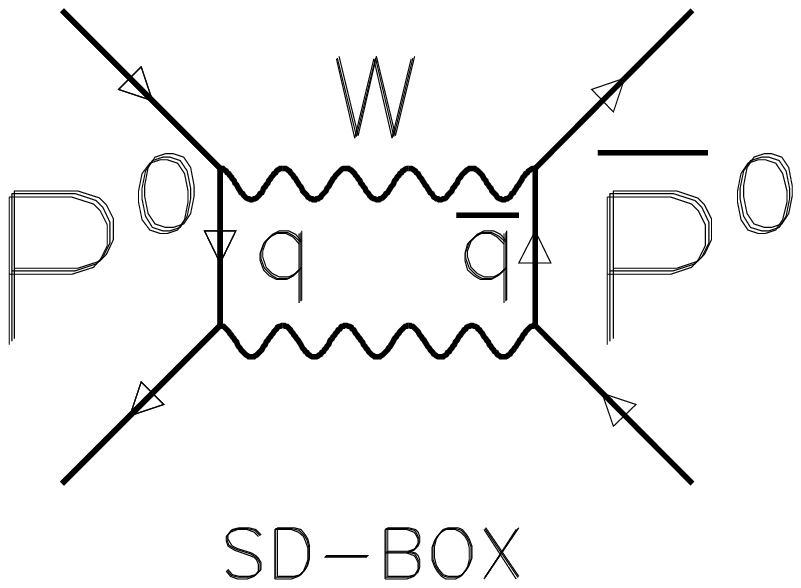}
 \includegraphics{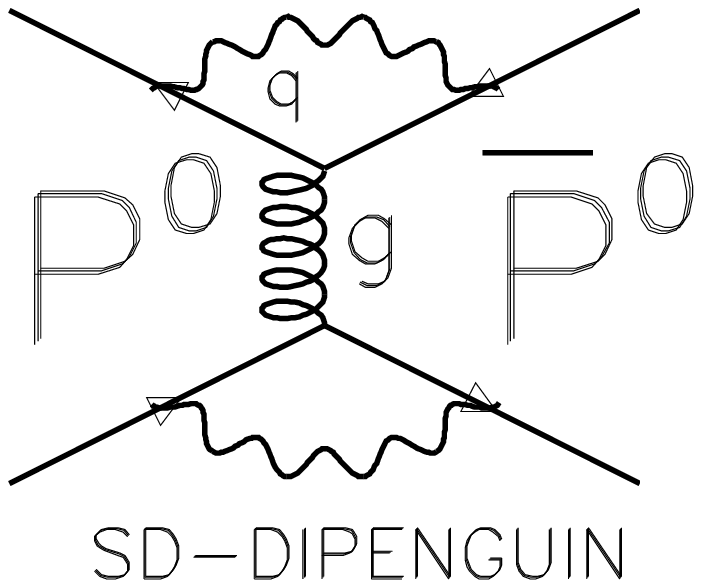}
 \includegraphics{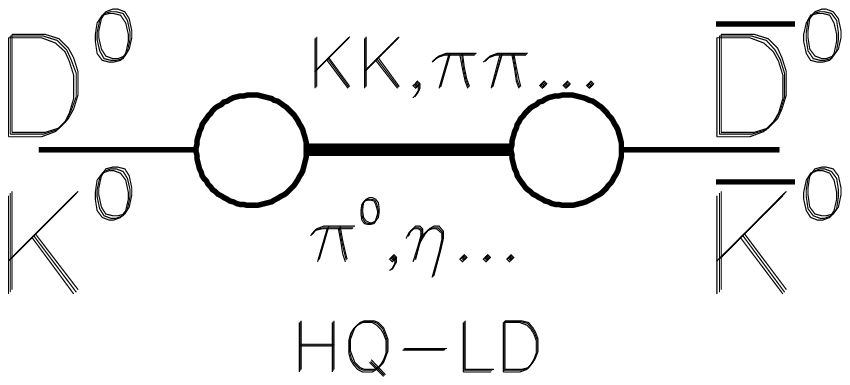}
 \caption{Box (proportional to $(m_q^2-m_u^2)/m_W^2$), penguin, and
 long-distance diagrams for mixing. \label{fig:box} }
\end{figure}
In the case of charm mesons,
%\cite{Golowich:1995xv}, 
because of the
Cabibbo-favored decay mechanism and the large phase space available for
their decay, decay widths are very similar $(y\ll 1)$,
and the time-integrated  ratio of mixed and nonmixed rates is
%\cite{Leader96}
\be
 r\equiv \frac{\Gamma(D^0\rarr\bar{D}^0\rarr \bar{f})}{\Gamma(D^0\rarr
 f)}  =  \qp2 \, \frac{x^2+y^2}{2}
\ee
\par
 Theoretical estimates of $x$ traditionally fall into two main categories,
 short distance 
 (SD) and heavy quark/long distance (HQ-LD): the former arise from the box
 diagram
%\cite{Golo4939} 
(Fig.\ref{fig:box}a),
 with GIM mechanism suppressing the
 charm case (Tab.\ref{tab:mix}) or the dipenguin diagram,
%\cite{Petrov:1997ch},
 the latter come from QCD diagrams
%\cite{Golo38}
 and
 final state interactions (FSI)
%\cite{Golo4939} 
such as
 rescattering of quarks with known intermediate light states
 (Fig.\ref{fig:box}c).
%\begin{wraptable}[9]{r}{4cm}
\begin{table}
 \caption{Box diagram contributions to mixing, and compilation of mixing
parameters (95\%\,CL )\cite{pdg2k}.
  \label{tab:mix}
 }
 \footnotesize
 \begin{center}
\begin{tabular}{lrcrr} \hline
 $P^0$              & $q$ & $SD/LD$    & $x$        &   $y$                  \\
\hline
% $K^0\,(d\bar s)$   & c & $SD\sim LD$ & $0.474\pm 0.001$ & $0.9965$
%\\ accorciata per fittarla
 $K^0\,(d\bar s)$   & c & $SD\sim LD$ & $0.474$ & $0.9965$       \\
 $D^0\,(c\bar u)$   & s & $SD\ll LD$  & $<0.03$          & $-0.06<y<0.01$ \\
% $B^0_d\,(d\bar b)$ & t & $SD\gg LD$  & $0.730\pm 0.029$ & $?$
%\\
 $B^0_d\,(d\bar b)$ & t & $SD\gg LD$  & $0.73$ & $?$            \\
 $B^0_s\,(s\bar b)$ & t & $SD\gg LD$  & $> 15.7$         & $<0.16$  \\
\hline
\end{tabular}
  \vfill
 \end{center}
%\end{wraptable}
\end{table}
 Recently, OPE-based approaches have also been proposed\cite{Bigi:2000wn}.
\par
 An important comment was made recently\cite{li97} on the possibility
 of measuring $y$ separately from $x$. Indeed, $x \neq 0$ means that mixing
 is genuinely produced by $\d0d0$ transitions (either SD or HQ-LD, or both),
 while $y \neq 0$ means that the fast-decaying component $D^0_1$ quickly
 disappears, leaving the slow-decaying component $D^0_2$ behind, which is a
 mixture of $D^0$ and $\bar{D}^0$. Infinite discussion is active on the
 extent to which the  three contributions are dominant: consensus seems to
 exist on the HQ--LD  being, in the case of charm mesons,  larger than the
 SD, and in  any case small.
 Standard Model  predictions are
%\cite{rpred}
\be
 x,y < 10^{-7}-10^{-3} \quad\quad r^{SM} <10^{-10}- 10^{-4}
\ee
 still somewhat below  the PDG2000 limit\cite{pdg2k} $r < 4.1\times 10^{-4}$.
 Any observation of $\d0d0$ mixing above the predicted level, once
  HQ--LD effects are understood, is a signal that new physics contributions
 are adding to the box diagrams. A recent compilation of predictions on $\d0d0$
 mixing is in Ref.~\cite{Nelson:1999fg}.
%\cite{Hewett:1996uc}.
Traditionally, $\d0d0$ mixing is searched for by  means of event-counting 
techniques, while advances in event statistics now allow studies of
the $y$ parameter.
\subsection{Wrong sign vs right sign counting }
 Mixing is searched for in the decay chains
\bea
  D^{*+} & \rarr & D^0 \pi^+      \nonumber                   \\ 
  D^0    &\Rightarrow & \bar{D}^0  \nonumber        \\
 \bar{D}^0 &  \rarr & K^+\pi^-,  K^+\pi^-\pi^+\pi^-, K^+\ell^- \bar{\nu}_\ell
\eea
 with the flavor of the neutral $D$ meson at production and at decay
  given by the sign of  $\pi^+$ and $K^-$
 respectively.
 In the case of a hadronic final state, life is complicated by pollution
 of the mixing by the Doubly-Cabibbo-Suppressed Decay $D^0 \rarr K^+\pi^-$,
 proportional to $\tan^4\theta_C$.
 The  measurable $r_{WS}$ -- the rate of wrong-sign events -- has therefore
 contributions\cite{Blaylock:1995ay}
 from DCSD, interference, and mixing
\bea
 r_{WS} & = & \Gamma(D^0\rarr f)/\Gamma(\bar{D}^0\rarr f)  \nonumber \\
 & = & \frac{e^{-\bar \Gamma t}}{4}|\langle f|H|D^0\rangle|^2_{CF}  \qp2
 (X+Yt+Zt^2) \nonumber \\
 X & \equiv & 4|\lambda|^2 \nonumber \\
 Y & \equiv & 2\Re(\lambda)\Delta\Gamma + 4 \Im(\lambda)  \Delta m  \nonumber\\
 Z & \equiv & (\Delta m)^2 +(\Delta\Gamma)^2/4 \nonumber \\
 \lambda & \equiv & \frac{p}{q}
              \frac{\langle f |H|D^0\rangle _{DCS}}
                   {\langle f  |H| \bar{D}^0\rangle _{CF}} 
\eea
 The X term (pure DCS) is characterized by  exponential decay time
 behavior, unlike the Z term (pure mixing), and this feature
 can in principle be used to suppress the DCS pollution. The Y
 (interference) term receives contributions from $\Im (\lambda)$, which can be
 nonzero if a) CPV is present, thus introducing a phase
 $\varphi\sim\arg(V_{cd}V_{ud}^*/V_{cs}V_{us}^*)$ in $p/q$;
 and/or b) a strong phase $\delta$ is present, due to different FSI
 in the DCS and mixed CF decays. By assuming CP conservation, i.e.,
 $|p/q|=1$,  defining
\be
 e^{i\varphi}\equiv\frac{p}{q} \quad\quad  e^{i\delta}\sqrt{r_{DCS}}\equiv
 \frac{\langle f|H|  D^0\rangle _{DCS}}{\langle f  |H|
 \bar{D}^0\rangle_{CF}}
\ee
 and
 measuring $t$ in units of $\bar{\Gamma}$ we can write
%\cite{li97}
  a simpler
 expression
 for $r_{WS}$
\be
 r_{WS} \propto e^{-t} [ r_{DCS} + t^2(r/2) +
 t\sqrt{2rr_{DCS}}\,\cos\phi ]
 \label{eq:ws}
\ee
 where the interference angle is given by $\phi=\arg(ix+y)-\varphi-\delta$.
 Equation \ref{eq:ws} shows how a meaningful quote of the $r$ result must
 specify which
 assumptions where made on the CPV and strong angles $\varphi$ and $\delta$.
 In particular, it was recently pointed
 out in\cite{Falk:1999ts}\cite{Bergmann:2000id} how the 
 possibility of a nonzero  strong phase $\delta$, which vanishes in the
 limit of
 unbroken SU(3) symmetry, should in fact be carefully taken into account in
 $\d0d0$ mixing, due to the experimentally known feature of SU(3) to be
 badly broken in D  decays.
 If one assumes CP invariance $(\varphi =0)$, then
\bea
 r_{WS} \propto e^{-t} \{ r_{DCS} + (r/2)t^2 +
 (y^\prime\sqrt{r_{DCS}})t \}  \\
 y^\prime \equiv y\cos\delta-x\sin\delta \quad x^\prime\equiv
 x\cos\delta+y\sin\delta
\eea
 The alternative option in counting techniques is the use of semileptonic
 final states $K \ell \nu$, which do not suffer from DCSD pollution but are
 harder experimentally.
\subsection{Lifetime difference measurements}
 The $y$ parameter can be determined  directly by measuring the
 lifetimes of  CP=+1 and CP=--1 final states -- such as  $K^+K^-$ and
 $\pi^+\pi^-$ (CP=+1),  $K_S\phi$  (CP=--1) -- and $K^-\pi^+$ (mixed CP), 
 assuming both CP conservation, i.e., that $D^0_1$ and
 $D^0_2$ are indeed CP eigenstates, and  that $K\pi$ is a mixed-CP
 eigenstate. We shall call such 
 parameter $y_{CP}$, and  if the above assumptions are verified
\begin{equation}
 y=y_{CP}=\frac{\tau(D\rarr K\pi)}{\tau(D\rarr KK)} -1
\end{equation}
 In principle, a measurement of $y$ would allow, along with
 an independent measurement of $r$, limits to be set on $x$. 
\subsection{CPV asymmetries}
 The assumption of negligible CPV in the charm system is important in the
 study of mixing.
 CPV occurs if the decay rate for a particle differs from the
 decay rate of its CP-conjugate particle \cite{Bigi:2000yz}.
 In charm meson decays (as well as in $K$ and $B$
 decays), two classes of CP violation exist: indirect and direct.
 In neutral charm meson decays, indirect CPV may arise
 due to $\d0d0$ mixing. In the case of direct violation,
 CP violating effects occur in a decay process only if the
 decay amplitude is the sum of two different parts, whose phases are made
 of a weak (CKM) and a strong contribution due to final state
 interactions  \cite{Bucc95}
\begin{equation}
 A\equiv ae^{i\delta_1}+be^{i\delta2}
\end{equation} 
The weak contributions to the phases
change sign when going to the CP--conjugate process, while the strong
ones do not. In singly Cabibbo-suppressed $D$ decays, penguin terms
in the effective Hamiltonian may provide the different phases of the
two weak amplitudes. The CP asymmetry will then be
\bea
 A_{CP} & \equiv &
   \frac{|A|^2-|\bar A|^2}{|A|^2+|\bar A|^2} \nonumber \\
        &  =     &
   \frac{2\Im(ab^*)\sin(\delta_2-\delta_1)}
        {|a|^2+|b|^2+2\Re(ab^*)\cos(\delta_2-\delta_1)}
\eea 
 Compared to the strange and bottom sectors, the SM predictions of CPV 
for charm decays are much smaller,
% \cite{Buccella,Bigi,Golden,Close},
making the charm sector a good place to test the SM and to
look for evidence of new physics. In the SM, direct CP
violating asymmetries in $D$ decays are predicted to be largest in
singly Cabibbo-suppressed decays, at most $10^{-3}$, and non-existent in
Cabibbo-favored and doubly Cabibbo-suppressed
decays\cite{Bigi:2000yz}. However, a CP asymmetry could occur
in the decay modes $D \to K_s \,\rm{n}\pi$ due to interference
between Cabibbo-favored and doubly Cabibbo-suppressed decays.
\begin{table*}[t]
\caption{Measured CP asymmetries ($\times 10^{-2}$). References to quoted
 results are in Ref.~\cite{Link:2000aw}}
\label{tab:asy}
 \footnotesize
\begin{tabular}{llll}
\hline
 Experiment                & $D^+ \to K^-K^+\pi^+$ & $D^0 \to K^-K^+$    &
                                                    $D^0  \to \pi^-\pi^+$ \\
\hline
  E687           & $-3.1 \pm 6.8$ & $+2.4 \pm 8.4$ &
              \\
  CLEO II        &                & $+8.0 \pm 6.1$ &
              \\
  E791           & $-1.4 \pm 2.9$ & $-1.0 \pm 4.9 \pm 1.2$ &
                                                    $-4.9 \pm 7.8 \pm 3.0$ \\
  FOCUS\cite{Link:2000aw} & $+0.6 \pm 1.1 \pm 0.5$ & $-0.1 \pm
   2.2 \pm 1.5$ &  
                                                    $ +4.8 \pm 3.9 \pm 2.5$ \\
\hline
\end{tabular}
\end{table*}
\section{FOCUS}
The data for this paper were collected in the wideband photoproduction
experiment FOCUS during the Fermilab 1996--1997 fixed-target
run. FOCUS is a considerably upgraded version of a previous experiment,
E687 \cite{Frabetti:1992au}. In FOCUS, a forward multi-particle
spectrometer is used to
measure the interactions of high-energy photons on a segmented BeO
target. We obtained a sample of over 1 million fully reconstructed
charm particles in the three major decay modes: $D^0 \rightarrow K^- \pi^+
,~K^- \pi^+ \pi^- \pi^+$ and $D^+ \rightarrow K^- \pi^+ \pi^+$.
\par
The FOCUS detector is a large-aperture, fixed-target spectrometer with
excellent vertexing, particle identification, and reconstruction
capabilities for photons and $\pi^0$'s. A photon beam is
derived from the bremsstrahlung of secondary electrons and positrons
with an $\approx 300$ GeV endpoint energy produced from the 800
GeV/$c$ Tevatron proton beam. The charged particles that emerge from
the target are tracked by two systems of silicon microvertex
detectors. The upstream system, consisting of four planes (two views in two
stations), is interleaved with the experimental target, while the
other system lies downstream of the target and consists of twelve
planes of microstrips arranged in three views. These detectors provide
high-resolution separation of primary (production) and secondary
(decay) vertices with an average proper time resolution of $\approx
30~ {\rm fs}$ for two-track vertices. The momentum of a charged particle
is determined by measuring its deflections in two analysis magnets of
opposite polarity with five stations of multiwire proportional
chambers. Three multicell threshold Cerenkov counters are used to
discriminate between pions, kaons, and protons, and complement the electron
identification provided by the em calorimetry. For each charged track, the
Cerenkov particle identification algorithm 
generates a set of $\chi^2$-like variables $W_i\equiv -2\log({\rm
likelihood})$ where $i$ ranges over the electron, pion, kaon, and proton
hypotheses. 
\section{SEARCH FOR CP-VIOLATING ASYMMETRIES}
 We have studied\cite{Link:2000aw} the Cabibbo-suppressed decay modes that
 have 
the largest combination of branching fraction and detection efficiency. For
this reason we selected the all-charged decay modes $D^+ \to K^-K^+\pi^+$,
$D^0 \to K^-K^+$, and $D^0 \to \pi^-\pi^+$ (charge
conjugate state implied, unless otherwise noted).
 In $D$ decays the charged $D$ is self-tagging and the neutral $D$ is tagged
as a $D^0$ or as a $\overline{D}^0$ by using the sign of the bachelor
pion in the $D^{*\pm}$ decay.  The CP asymmetry parameter measures the
direct CP asymmetry in the case of $D^+$ and the combined direct and indirect
CP asymmetries for $D^0$.
% \cite{Palmer}.
\par
 Before searching for CP asymmetry we must account for
differences, at the production level, between $D$ and $\overline{D}$
in photoproduction (the hadronization process, in the presence
of remnant quarks from the nucleon, gives rise to production
asymmetries\cite{Gardner}). This is done by normalizing to the
Cabibbo-favored modes $D^0 \to K^-\pi^+$ and $D^+ \to K^-\pi^+\pi^+$,
with the additional benefit that most of the corrections due to inefficiencies
cancel out, reducing systematic uncertainties. An implicit assumption is
that there is no measurable CPV in the Cabibbo-favored decays.
The CP asymmetry can be written as:
\begin{equation}
 A_{CP} = [\eta(D)-\eta(\overline{D})]/[\eta(D)+\eta(\overline{D})]
\end{equation}
 where $\eta$ is (considering, for example, the decay mode $D^0 \to K^-K^+$):
\begin{displaymath}
 \eta(D) = \frac{N(D^0 \rightarrow K^-K^+)}{N(D^0 \rightarrow K^-\pi^+)}
           \frac{\epsilon(D^0 \rightarrow K^-\pi^+)}{\epsilon(D^0
           \rightarrow K^ 
-K^+)}
\end{displaymath}
where $N(D^0 \rightarrow K^-K^+)$ is the number of reconstructed candidate
decays and $\epsilon(D^0 \rightarrow K^-K^+)$ is the efficiency obtained
from montecarlo simulations.
\par
Table~\ref{tab:asy} shows the asymmetry numbers obtained by FOCUS, compared
to previous published asymmetry measurements. The statistical error
of the neutral decay channel is not as good as the charged one, since 
$D^*$ tagging is necessary to determine the flavor of the parent
$D^0$. There is no clear evidence for CPV in our measurements, which
correspond to new limits two-three times better than the previous
measurements by 
E791.
\section{LIFETIME DIFFERENCES AND $y_{CP}$ MIXING PARAMETER}
We have studied\cite{Link:2000cu} the  $y_{CP}$ mixing parameter by
measuring the difference in 
lifetimes for the $D^0$ decays: $D^0 \rightarrow K^-\pi^+$ and $D^0
\rightarrow K^-K^+$.
\subsection{Event selection}
\begin{figure}[ht!]
%
%\begin{center}
%	\includegraphics[height=1.5in]{kpi_wk4_new.ps}
%	\includegraphics[height=1.5in]{kk_wk1_new.ps}
%	\includegraphics[height=1.5in]{kk_wk4_new.ps}
%\end{center}
%
%
 \vspace{4.5in}
 \includegraphics{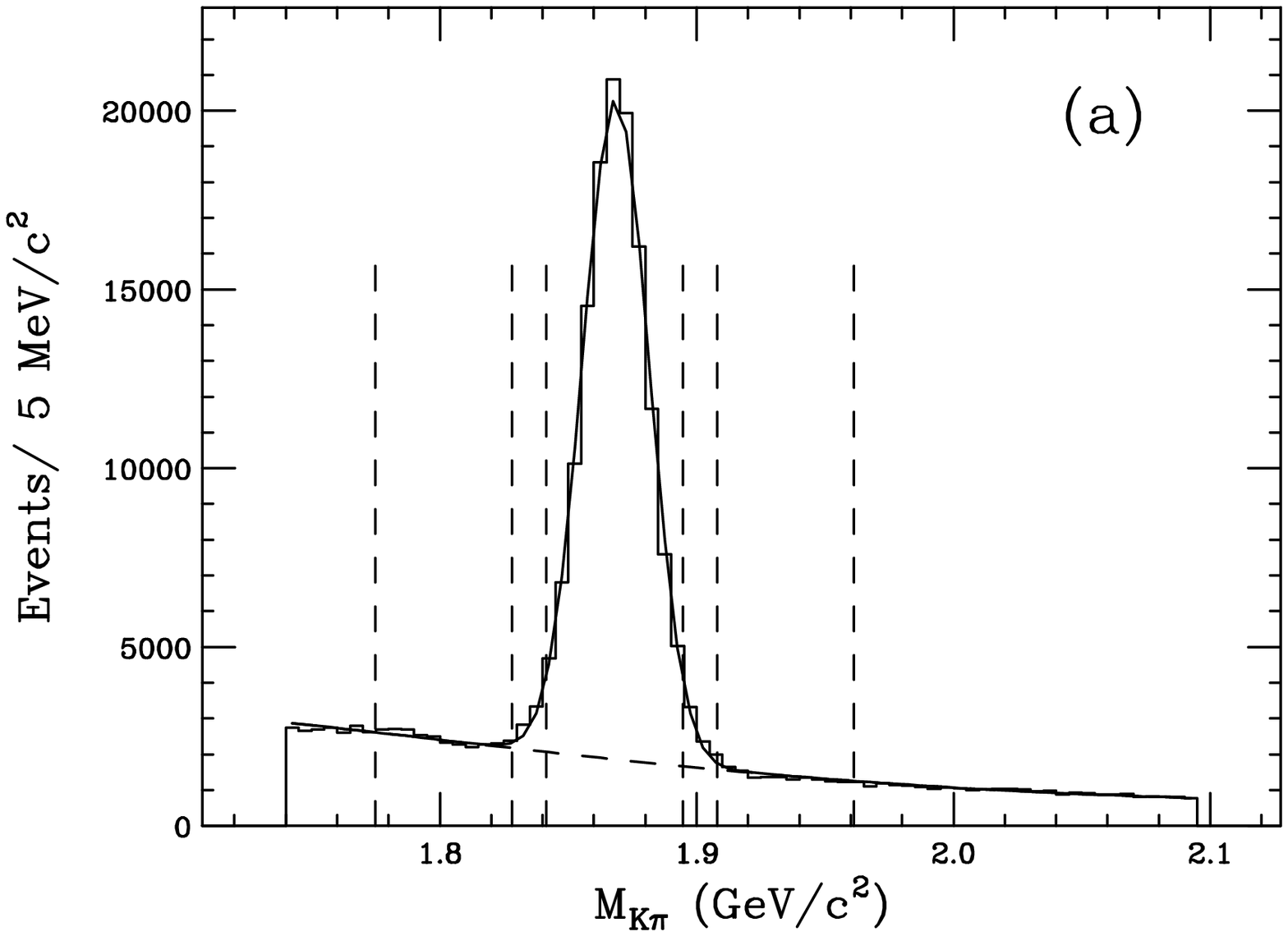}
 \includegraphics{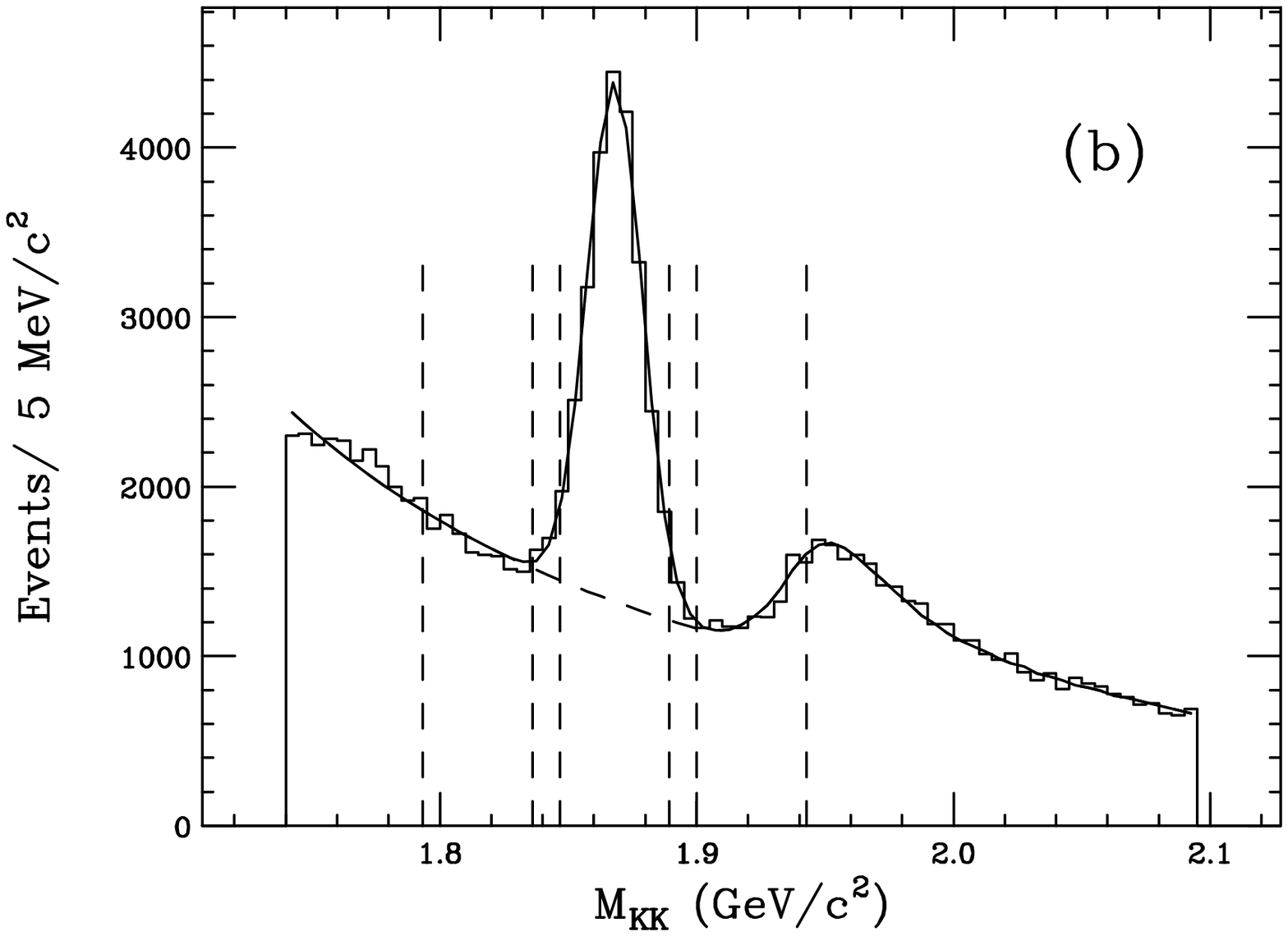}
 \includegraphics{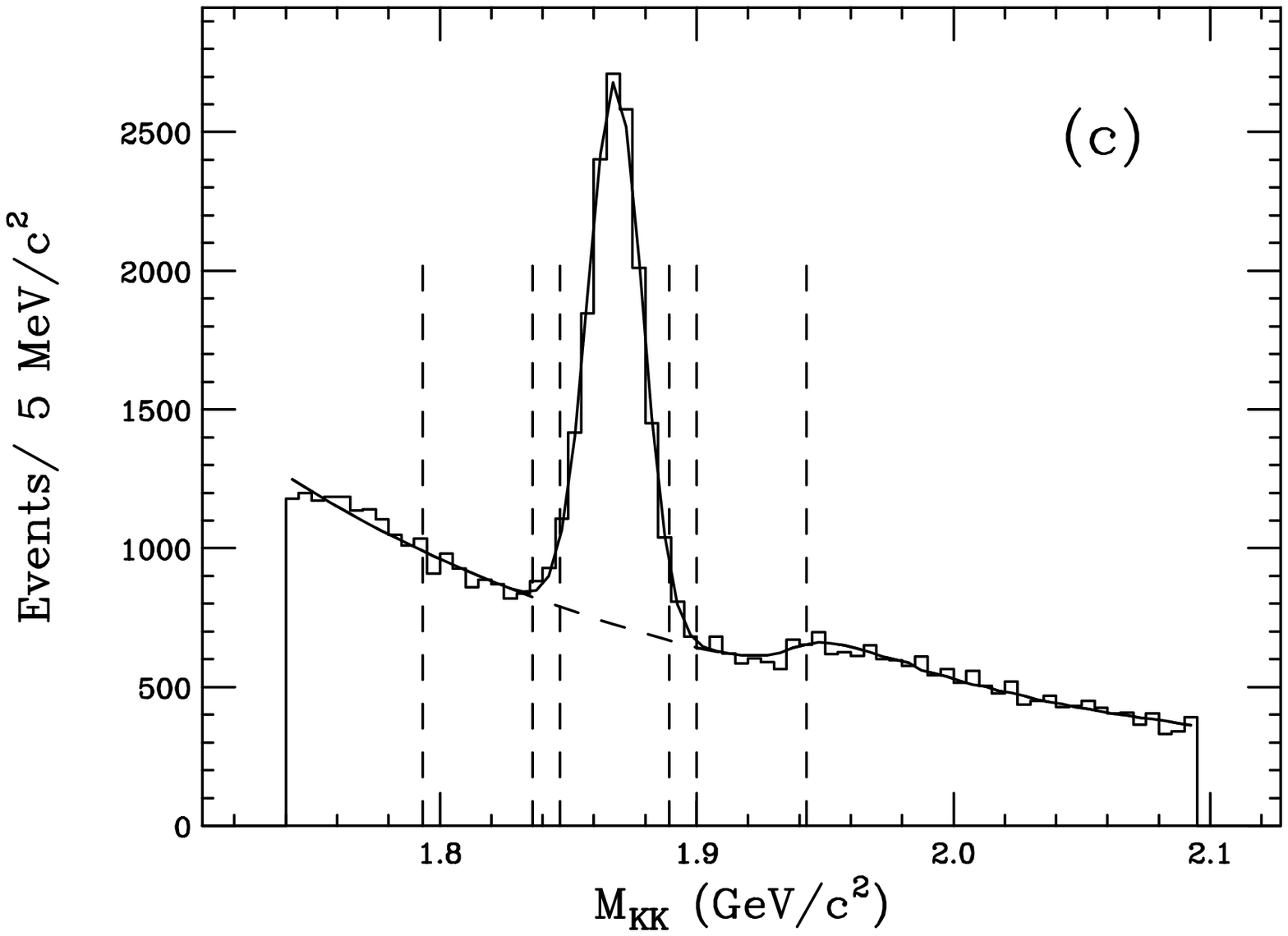}
\caption{ (a) Signal for $D^0 \rightarrow K^- \pi^+$ with a detachment
cut of $\ell/\sigma > 5$ and $W_\pi - W_K > 4$. The yield is 119\,738
$K^- \pi^+$ signal events. (b,c) Signals for $D^0 \rightarrow K^-
K^+$ with a detachment cut of $\ell/\sigma > 5$. The reflection in the
background at higher masses is due to contamination from misidentified
$D^0 \rightarrow K^- \pi^+$.  (b) Requiring $W_\pi - W_K > 1$, we
obtain a yield of 16\,532 $K^- K^+$ signal events.  (c) Requiring $W_\pi
- W_K > 4$, we obtain a yield of 10\,331 $K^- K^+$ signal events.  The
vertical dashed lines indicate the signal and sideband regions used
for the lifetime and $y_{\rm CP}$ fits.}
\label{fig:signals}
\end{figure}

The cuts used to obtain a clean signal were designed to produce a nearly
flat efficiency in reduced proper time $t^\prime \equiv (\ell - N
\sigma)/(\gamma\beta c )$,
which is defined as the proper time subtracted event-by-event by the
minimum amount of detachment required between primary and secondary
vertices. Our quoted 
result was based on requiring a minimum $\sigma_\ell$ detachment and kaon
hypothesis over pion hypothesis in Cerenkov response favored for kaon
candidates. Then we either require a $D^*$ tag, or a set of
more stringent cuts, such as more stringent Cerenkov requirements on kaons
and pions, momenta of decay particles balancing each other, 
primary vertex inside the target material, and resolution of proper time
less than 60~fs. The $D^*$ tagged sample has a better signal-to-noise ratio,
while the inclusive sample accommodates larger sample size. From the
combination of two samples, we obtain 119\,738 $D\rarr K\pi$ and 10\,331
$D\rarr KK$ events (Fig.~\ref{fig:signals}).
\par
\subsection{Fitting technique}
\begin{figure}[h!]
%\begin{center}
%	\includegraphics[height=1.5in]{newft.eps}
%\end{center}	
%
 \vspace{1.3in}
 \includegraphics{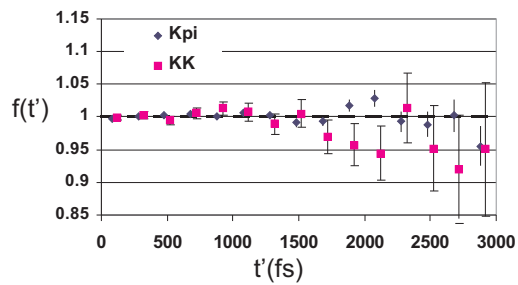}
\caption{ Montecarlo correction factors for $D^0 \rightarrow K^-
\pi^+ ~{\rm and}~ K^- K^+$ for $\ell/\sigma > 5$ and $W_\pi -
W_K > 4$. We have offset the $K^- K^+$ points slightly for clarity and
have given them ``flats'' on their error bars.  Montecarlo corrections
are rather slight with these cuts and the corrections for $D^0
\rightarrow K^- \pi^+$ are the same within errors as those for $D^0
\rightarrow K^- K^+$.  }
\label{fig:ft}
\end{figure}

The $D^0\rarr K^-K^+$ sample is characterized by a prominent reflection
background coming from misidentified $D^0\rarr K^-\pi^+$ decays
(Fig.~\ref{fig:signals}). 
We accommodate the reflection effect by using a modified version of the mass
sideband subtraction  fitting technique used in the E687
experiment\cite{Frabetti:1992au}. The amount of  $D^0\rarr K^-\pi^+$
reflection 
is obtained 
by a mass fit to the $K^-K^+$ sample and the shape of the reflection is
deduced from a high-statistics montecarlo sample. We assume that the time
evolution of the reflection is described by the lifetime of $D^0\rarr
K^-\pi^+$ and we fit the reduced proper time distributions of the
$D^0\rarr K^-\pi^+$ and $D^0\rarr K^-K^+$ samples at the same time. The fit
parameters are the $D\rarr K\pi$ lifetime, the lifetime asymmetry $y_{CP}$,
and the number of background events under each 
$D^0\rarr K^-\pi^+$ and $D^0\rarr K^-K^+$ signal region. The signal
contributions for the $D^0\rarr K^-\pi^+$, $D^0\rarr K^-K^+$ and the
reflection from the misidentified $D^0\rarr K^-\pi^+$ in the reduced proper
time histograms are described by  a term
\begin{equation}
 f(t^\prime)\, \exp(-t^\prime/\tau)
\end{equation}
in the fit likelihood function. The function $f(t^\prime)$, determined from
montecarlo,  covers any 
deviation of the reduced proper time distribution from a pure exponential
due to acceptance (Fig.~\ref{fig:ft}).
\par
\begin{figure}[h!]
%\begin{center}
%	\includegraphics[height=1.5in]{logtime.ps}
%\end{center}	
%
 \vspace{1.5in}
 \includegraphics{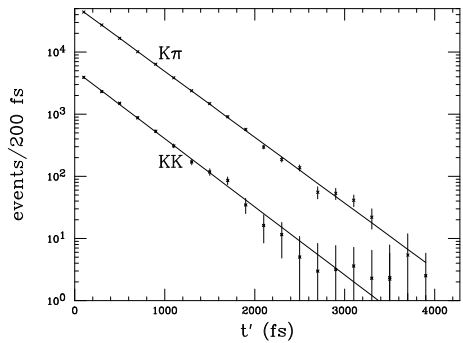}
\caption{ Signal versus reduced proper time for $D^0 \rightarrow K^-
\pi^+~ ~{\rm and}~~K^- K^+$ requiring $W_\pi - W_K > 4$ and
$\ell/\sigma > 5$. The fit is over 20 bins of 200 fs bin width. The
data is background subtracted and includes the (very small) montecarlo 
correction.  }
\label{fig:life}
\end{figure}
 The background yield parameters are either left floating, or fixed to the
number of events in mass sidebands using a Poisson penalty term in the fit
likelihood function. We choose 200~fs as the bin size of the reduced proper
time, which is large compared to our proper time resolution. Twenty bins
are used in the fit (Fig.~\ref{fig:life}).
\par
The systematic errors are estimated by changing the selection cuts and by
trying different fitting methods. We tested the Cerenkov identification
hypothesis for kaon candidates and the minimum detachment required between
primary and secondary vertices. The former affects the level of reflection
backgrounds, the latter affects the amount of non-charm
backgrounds. We tried different numbers of histogram bins and two options of
background handling, as stated in the previous subsection. 
Other variations of selection and fitting yielded results consistent with our
number. We obtained 
\bea 
 y_{CP} = (3.42 \pm 1.39 \pm 0.74)\%  \\
 \tau(D\rarr K\pi) = 409.2 \pm 1.3 {\rm fs}
\eea 
Our result on $\tau(D\rarr K\pi)$ has a statistical error only. Detailed
systematics 
studies, including absolute distance scale, are needed to obtain the final
number. 
\section{PRELIMINARY MEASUREMENT OF 
         $\Gamma(D^{*+}\rightarrow \pi^+ (K^+\pi^-))/
          \Gamma(D^{*+}\rightarrow \pi^+ (K^-\pi^+))$}
As described in \S~2.1, the $D^0$ can decay to $K^+\pi^-$ through two
physical processes: 1) by a DCS decay or 2) by mixing to $\bar{D}^0$
followed by the CF decay to $K^+\pi^-$. The Standard Model predicts a DCS
to CF branching ratio $(r_{DCS})$ of the order $\tan^4\theta_c\simeq
0.25\%$. The Standard Model predictions for $\d0d0$ mixing rate were
discussed in  \S~2. We also discussed how CP violation can
cause rate asymmetries for both mixing and DCS decays. In this
analysis\cite{Link00}  we
ignore possible CPV effects which, in the SM, are expected to be small
compared to the current experimental sensitivity.
\par
The selection algorithms and analysis cuts are identical for both the mode
under study $D^0\rarr K^+\pi^-$ and the normalizing mode $D^0\rarr
K^-\pi^+$. To separate these modes we tag the flavour of the neutral $D$
meson via the decay $D^{*+}\rarr D^0\tilde\pi^+$. 
In describing the event selection procedure we shall refer to the events
consistent with the (dominant) CF process $D^{*+}\rightarrow \tilde\pi^+
(K^-\pi^+)$ as right-sign
(RS) tagged, and to the events consistent with DCS decay or mixing as
wrong-sign (WS) tagged. 
\begin{figure}[ht!]
%
%  \vspace{1cm}
%   \begin{center}
%     \epsfbox{draft10-23p6.eps}
%   \end{center}
 \vspace{5.5cm}
% \special{psfile=draft10-23p6.eps vscale=110 hscale=110 voffset=-20 hoffset=15
%  angle=0}
 \includegraphics{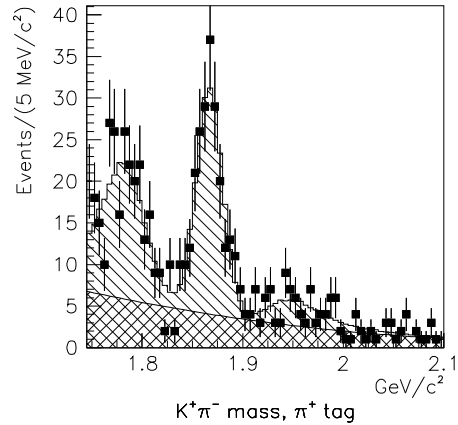}
  \caption{An example fit to the $K\pi$ mass from inside the DCS-like
signal region $(146\,{\rm MeV} < \Delta m < 147\,{\rm MeV})$. Data
points are the squares with error bars,  fit to data points is shown as the
slashed histogram, and the polynomial background fit function is cross-hatched.
   }
   \label{fig:d*slice} 
\end{figure}
\subsection{Event selection}
Candidate events consist of a pair of oppositely charged tracks that form a
vertex and have a $K\pi$ invariant mass between 1.7 and 2.1
GeV/$c^2$. These $D^0$ candidates are used as a seed to search for a
suitable production vertex consisting of at least two other charged
tracks. The production vertex is required to be isolated from both $D$
candidate daughter tracks, the production vertex is also required to be in
target material.  The production and decay vertices have to be well
separated, and both primary and secondary have to be formed with a good
confidence level. 
To remove background that results from a high-momentum track paring with
a random low-momentum track to form a $D$ candidate, we apply a
momentum-dependent 
cut is applied that removes highly asymmetrical $K\pi$ pairs.
\par
Doubly misidentified $K^+\pi^-$ pairs from $D^0$ decays form a broad peak
directly under the $D^0$ signal in $K^+\pi^-$ and  a narrow peak in the $D^*-D$
mass difference  signal region. The mass difference background is
indistinguishable  from the real WS tagged signal. To eliminate this
background, the $K\pi$ invariant mass is computed with the kaon and pion
particle hypotheses swapped. Any candidate whose swapped mass is within
$\pm 4\sigma$ of the $D^0$ mass is subjected to a cut on the sum of the
$K\pi$ separations $(W_\pi-W_K)$ for both tracks. 
Finally, all tracks in the production vertex are tested as potential
$\tilde\pi$ candidates, and are accepted if within a $\pm 50\,{\rm MeV}/c^2$
window of the nominal $D^*-D^0$ mass difference, and if they satisfy a loose
Cerenkov cut. 
\begin{figure*}[ht!]
  \vspace{3.3cm}
 \includegraphics{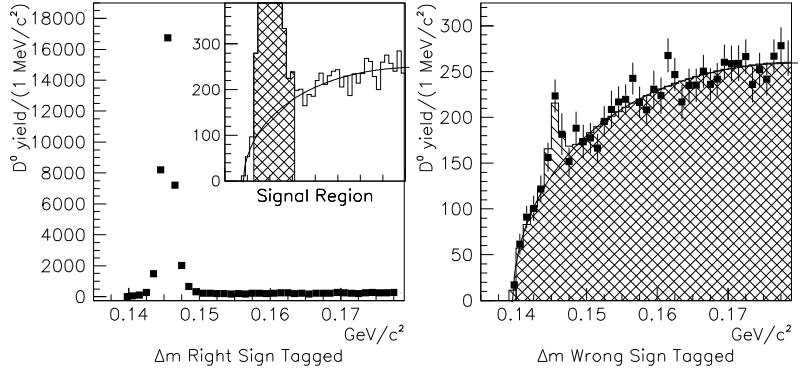}
%  \epsfbox{draft10-23p7.eps}
  \caption{(Left) The RS mass difference distribution. In the inset 
the signal 
region used as the WS model is cross-hatched and the solid curve is the fit
background shape. (Right) The WS mass difference distribution. Squares with
error bars are the fitted $D^0$ yields, and histogram is the
fit. Background fit is cross-hatched and signal fit is slashed.
    }
  \label{fig:kpisb}
\end{figure*}
\subsection{Fitting technique}
Reflections from partially reconstructed and/or misidentified $D^0$ decays
with a real $\tilde\pi$ can contribute significantly to the WS signal. The
measurement method adopted allows one to deal with reflections and feed
downs from all known $D^0$ decays ($D^0\rarr K^+K^-$ and $D^0\rarr
\pi^+\pi^-$ reconstructed as $D^0\rarr K^+\pi^-$, $D^0\rarr K^+\pi^-\pi^0$
partially reconstructed, $D^0\rarr K^-\pi^+\pi^0$ and $D^0\rarr
K^-\ell^+\nu$ partially reconstructed and doubly misidentified).
To deal with these backgrounds we isolate the true $D^0\rarr K^+\pi^-$
decays by fitting the $K\pi$ invariant mass distribution. 
The $K\pi$ invariant mass distribution is generated by splitting the RS
and WS tagged samples into 1-MeV-wide bands in the $D^*-D$ mass differences
(Fig.\ref{fig:d*slice}). Reflections
($KK$ and $\pi\pi$) are fit to montecarlo line shapes, the unstructured
background is fit to a polynomial, and the $D^0$ signal is fit to a Gaussian.
The fitted $D^0\rarr K\pi$ yields are plotted as a function of the
appropriate mass difference bins (Fig.\ref{fig:kpisb}).  
By fitting the $D^0$ in this way we generate a mass difference distribution
that has only true $D^0\rarr K\pi$ events: the signal is true $D^0\rarr
K\pi$ events with a true $\tilde\pi$ tag, and the background is true
$D^0\rarr K\pi$ with a random $\tilde\pi$ tag. The random tagged events
form a smooth threshold background which is parametrized by the function
\begin{equation}
 f(\Delta m)\equiv \alpha (\Delta m - m_\pi)^{1/2} +
                   \beta(\Delta m -m_\pi)^{3/2}  
\end{equation}
where $\alpha$ and $\beta$ are free fit parameters.
The RS signal is used in the fit as a shape model for the WS
signal. At each fit iteration, the RS background -- fit outside the signal
region -- is subtracted from the RS distribution. The final fit
parameter is a scale factor used to match the background-subtracted RS
signal to the WS signal. In the WS $D^*$ signal region the total fit
function is the sum of the scaled RS signal and the WS background
parametrization. This signal scale factor is the WS to RS branching
ratio. The fit is shown in Fig.\ref{fig:kpisb}, and we obtain a preliminary WS
to RS branching ratio (statistical error only)
\begin{equation}
\frac{\Gamma(D^{*+}\rightarrow \pi^+ (K^+\pi^-))}
     {\Gamma(D^{*+}\rightarrow \pi^+(K^-\pi^+))} = (0.482\pm0.093)\%
\end{equation}
From the fit we find $35901\pm196$ RS events corresponding to a WS
equivalent yield of $173\pm34$. The result above is preliminary, only
the statistical error is quoted. 
\begin{table}
 \caption{Recent results on $\d0d0$ mixing parameters. The CLEO limit on
$y^\prime$ assumes $x^\prime=0$.}
  \label{tab:ycpcom}
 \footnotesize
\begin{tabular}{ll} 
\hline
 E791~\cite{Aitala:1999nh}  & $y_{CP}=(0.8\pm 2.9 \pm 1.0) \%$     \\
 CLEO~II.V\cite{Godang:2000yd} & $(-5.8<y^\prime<1.0)\%\quad 95\%\,CL$  \\
 BELLE prelim.\cite{Tanaka:2000xw} & $y_{CP}=(1.0^{+3.8}_{-3.5}{}^{+1.1}_{-2.1})\%$ \\
 FOCUS\cite{Link:2000cu} & $y_{CP}=(3.42\pm 1.39 \pm 0.74) \%$
\\ 
\hline
\end{tabular}
  \vfill
\end{table}
\section{CONCLUSIONS}
We have presented a measurement of the $D^0\rarr KK$ and $D^0\rarr K\pi$
lifetime indicating that the $KK$ eigenstate has a shorter lifetime
than $K\pi$
\begin{equation}
 y_{CP} = (3.42\pm 1.39 \pm 0.74) \%
\end{equation}
Our $y_{CP}$ value is compared in Tab.~\ref{tab:ycpcom} with E791 and
BELLE. Also reported is the recent CLEO measurement of $y^\prime$ 
by studying the time evolution of WS hadronic decays (see \S~2.1).
Comparison with  the CLEO measurement is not clear because of lack of
information on the strong phase
$\delta$ \cite{Falk:1999ts}\cite{Bergmann:2000id}, and 
any comparison of  the $y_{CP}$ and the
$y^\prime$ into one $y$ parameter should be taken {\it cum grano
salis}. \cite{therev}
\par
We have also presented new limits
on CPV asymmetries for Cabibbo-suppressed decays such as $KK\pi$, $KK$ and
$\pi\pi$. All results are consistent with zero, with errors at the percent
level. 
\par
Finally, we showed a preliminary measurement of the WS to RS
branching ratio, which, 
assuming no mixing, corresponds to (statistical error only)
\begin{equation}
r_{DCS} = (0.482\pm 0.093)\%
\end{equation}
This preliminary result preludes  a full-blown lifetime
analysis which soon will 
provide the  FOCUS measurement of the $r^\prime$ mixing parameter.
\par
We wish to acknowledge the assistance of the staffs of the Fermi National
Accelerator Laboratory, the INFN of Italy, and the physics departments
of the collaborating institutions. This research was supported in part
by the U.~S.~A.  National Science Foundation, the U.~S.~A. Department of
Energy, the Italian Istituto Nazionale di Fisica Nucleare and
Ministero dell'Universit\`a e della Ricerca Scientifica e Tecnologica,
the Brazilian Conselho Nacional de Desenvolvimento Cient\'{\i}fico e
Tecnol\'ogico, CONACyT-M\'exico, the Korean Ministry of Education, and
the Korean Science and Engineering Foundation.
\par

\end{document}